# Synthesis and transport properties of ternary type-I Si clathrate $K_8Al_7Si_{39}$

Shiva Kumar Singh[1], Takashi Mochiku[1], Soshi Ibuka[2], Yukihiro Isoda[1], Akinori Hoshikawa[3], Toru Ishigaki[3], and Motoharu Imai[1]

[1] National Institute for Materials Science, Tsukuba, Ibaraki 305-0047, Japan

[2] Institute of Materials Structure Science, High Energy Accelerator Research Organization, Tokai, Ibaraki 319-1106, Japan

[3] Frontier Research Center for Applied Atomic Science, Ibaraki University, Tokai, Ibaraki 319-1106, Japan

*E-mail:* IMAI.Motoharu@nims.go.jp

## Abstract

A ternary type-I Si clathrate, $K_8Al_xSi_{46-x}$, which is a candidate functional material composed of abundant non-toxic elements, was synthesized and its transport properties were investigated at temperatures ranging from 10 to 320 K. The synthesized compound is confirmed to be the ternary type-I Si clathrate $K_8Al_7Si_{39}$ with a lattice parameter of $a = 10.442$ Å using neutron powder diffractometry and inductively coupled plasma optical emission spectrometry. Electrical resistivity and Hall coefficient measurements revealed that $K_8Al_7Si_{39}$ is a metal with electrons as the dominant carriers at a density of approximately $1\times10^{27}$ /m$^3$. The value of Seebeck coefficient for $K_8Al_7Si_{39}$ is negative and its absolute value increases with the temperature. The temperature dependence of the thermal conductivity is similar to that for a crystalline solid. The dimensionless figure of merit is approximately 0.01 at 300 K, which is comparable to that for other ternary Si clathrates.

# 1. Introduction

Ternary type-I 14-group clathrates, of which the general chemical formula is $M_8E_xX_{46-x}$ (M: alkaline, alkaline-earth elements; E: group 12, 13, and transition metal elements; X: Si, Ge, and Sn), have been extensively studied as thermoelectric materials [1-7] since Nolas et al. reported the ternary type-I Ge clathrate $Sr_8Ga_xGe_{46-x}$ with low glasslike thermal conductivity [8]. $M_8E_xX_{46-x}$ has a characteristic structure, the $Na_8Si_{46}$-type (cubic, space group: $P\bar{m}3n$, No 223, Z=1) [9], in which the E and X atoms form a polyhedral cage framework that can encapsulate guest M atoms. $M_8E_xX_{46-x}$ is considered to be a material that realizes the phonon-glass electron-crystal (PGEC) concept, which encompasses the electrical properties of crystalline materials and the thermal properties of amorphous or glasslike solids [10]. In $M_8E_xX_{46-x}$, the guest M atoms vibrate locally inside the cages, which is referred to as rattling. The interaction between the rattling mode of guest atoms and the acoustic mode of the cage framework is considered to cause the low thermal conductivity [11]. In $Ba_8Ga_xGe_{46-x}$, a type-I Ge clathrate, a dimensionless figure of merit *ZT* close to 1 [12-14] and carrier-type tuning by changing the Ga content [15,16] have been reported, demonstrating that the Ge clathrate is promising for practical thermoelectric applications. When the sustainability of raw material supplies is taken into account, Si-based materials have an advantage over Ge-based materials because Si is more abundant in the Earth's crust. Although the thermoelectric properties have been studied for ternary Si clathrates with encapsulated alkaline-earth metals, such as $Ba_8Ga_{16}Si_{30}$ [17] and $Ba_8Al_{14}Si_{31}$ [18], the *ZT* value for the ternary Si clathrates was found to be much less than 1. Therefore, it is necessary to find ternary Si clathrates with superior *ZT* values.

The physical properties of ternary type-I Si clathrates with encapsulated alkaline atoms have been investigated recently. Calculation of the electronic structures of $A_8Ga_8Si_{38}$ (A=Na, K, Rb, and Cs) and $K_8Al_8Si_{38}$ reveals that these are indirect gap semiconductors with energy gaps ($E_g$) that range from 0.20 eV below to 0.24 eV above the $E_g$ value for elemental Si with the diamond-type structure [19,20]. Elemental Si has an $E_g$ of 1.1 eV; therefore, the $E_g$ values of these clathrates range from 0.90 to 1.34 eV. These $E_g$ values are suitable for solar cell applications because a single-junction solar cell exhibits maximum conversion efficiency at a photon energy of 1.4 eV [21]. Recently, $K_8Ga_8Si_{38}$ was experimentally revealed to be a semiconductor [19] with an $E_g$ of 1.2 eV [22]. As the next step, we attempted to replace Ga with Al in consideration of the sustainability of raw material supplies and toxicity issues. During this trial, two reports on $K_8Al_xSi_{46-x}$ were published. One reported that $K_8Al_8Si_{38}$ is a semiconductor with an energy gap of approximately 1 eV on the basis of experimental and computational results, whereby it was concluded that $K_8Al_8Si_{38}$ is a promising material for solar cell applications [23]. Another demonstrated that $K_8Al_8Si_{38}$ has a thermal conductivity of approximately 1 $WK^{-1}m^{-1}$ by first-principles calculations [24]. However, there have been no experimental reports on the transport properties of $K_8Al_xSi_{46-x}$.

In this study, the ternary Si clathrate $K_8Al_7Si_{39}$ was synthesized, its crystal structure was characterized, and its transport properties such as the electrical resistivity $\rho$, Hall coefficient $R_H$, Seebeck coefficient $S$, thermal conductivity $\kappa$, and $ZT$ were investigated at temperatures in the range from 10 to 320 K.

## 2. Experimental procedure
### 2.1 Sample synthesis

A 10:8:38 molar mixture of K (98%, Sigma-Aldrich, pieces), Al (99.99%, Furuuchi Chemicals, powder), and Si (99.9999%, Furuuchi Chemicals, powder) was used as the starting material. After mixing the Al and Si powder, K pieces were then added. The starting material was loaded into a boron nitride (BN) capsule with a lid, which was then sealed in a stainless-steel (SS) tube with SS fittings. Weighing and mixing of the materials, and sealing of the BN tube in the SS tube were performed under an argon atmosphere in a glove box ($H_2O$ <1 ppm and $O_2$ <1 ppm). The SS tube was heated to 1353 K and held for 12 h, and subsequently cooled naturally. The resultant sample includes Si as an impurity phase; therefore, it was ground and heated at the same temperature for 24 h in a BN capsule sealed in an SS tube to obtain a homogeneous sample. The obtained sample was ground well again and washed using dilute HCl to remove any impurity phase after ultrasonication in ethanol. The final product was then obtained after washing with water. Details on how to determine the appropriate synthesis conditions have been described elsewhere [25].

Sintered samples for transport measurements were also prepared from the as-synthesized powder by spark plasma sintering (SPS), where the powder was sandwiched between graphite sheets in a graphite cylinder and sintered in an Ar atmosphere using a tungsten carbide (WC) die and WC pistons at 550 MPa and 923 K for 30 min.

**2.2 Characterization of as-synthesized samples**

The as-synthesized samples were confirmed to be single-phase clathrate by powder X-ray diffraction (XRD) measurement with Cu Kα radiation. The chemical compositions of the powder samples were determined by inductively coupled plasma optical emission spectrometry (ICP-OES). Samples for ICP-OES measurements were pretreated as follows. A powder sample (0.1 g) was decomposed in 10 mL of 7 M nitric acid and 10 mL of 6 M hydrochloric acid with heating in a 100 mL polytetrafluoroethylene (PTFE) beaker. The residue was filtered using filter paper and the filtrate was stored in a PTFE beaker covered with a PTFE plate (STEP 1). The residue and the used filter paper were transferred into a nickel crucible and ashed. After sodium carbonate (2 g) and sodium peroxide (3 g) were weighed into the crucible, the crucible was heated to 973 K in a furnace and held for 10 min and then cooled outside the furnace. The crucible with the molten salt was transferred into a 200 mL PTFE beaker and the molten salt was dissolved in 100 mL of 4 M hydrochloric acid. The crucible was removed from the PTFE beaker and rinsed with high-purity water. The solution was mixed with the filtrate from STEP 1 and a spike solution containing yttrium was added as an internal standard for ICP-OES measurement. The mixed solution was transferred into a 250 mL volumetric flask and diluted to 250 mL with high-purity water before measurement.

The crystal structure of the clathrate was determined by neutron powder diffractometry. The neutron powder diffraction (NPD) data were obtained with a time-of-flight (TOF) high-throughput powder diffractometer (iMATERIA, Ibaraki Materials Design Diffractometer) [26] at the Materials and Life Science Experimental Facility (MLF) of J-PARC. The high-resolution bank, which covers $0.09 < d$ (Å) $< 5.0$ with the resolution $\Delta d/d \sim 0.16\%$ was utilized. The sample was contained in a cylindrical vanadium cell (6 mm diameter and 1 mm length) and placed in a vacuum chamber. The NPD intensity data were collected at room temperature.

Rietveld refinement [27] was performed on the observed NPD pattern using the software FULLPROF [28]. The $Na_8Si_{46}$-type structure [9] was adopted as a structural model under the following constraints: K occupies the Na sites of $Na_8Si_{46}$, and Al and Si randomly occupy the Si sites. Al and Si at the same Wyckoff site share positional coordinates $x$, $y$, $z$ and an isotropic temperature parameter $B_{iso}$. The site occupancy is 1 in total at each site. The peak-shape function is a convolution of a pseudo-Voigt function with a pair of back-to-back exponentials [29]. The background intensity was provided with 14 user-defined intensity points with linear interpolation for intermediate values.

## 2.3 Characterization of sintered samples

SPS samples were characterized by powder XRD measurement (Cu Kα radiation) and electron-probe microanalysis (EPMA). The electron-probe microanalyzer was operated at an accelerating voltage of 15 kV and a beam current of 50 nA. $KNbO_3$, $Al_2O_3$, and Si were used as standard materials. For EPMA, the samples were mounted in a resin and polished using an oil-based diamond slurry.

The density of the SPS samples was measured using the Archimedes principle by measuring the weight of samples in air and water.

## 2.4 Electrical transport measurements

The electrical resistivity, Seebeck coefficient, and thermal conductivity of the SPS samples were measured at temperatures ranging from 10 to 320 K using a physical property measurement system (PPMS; Quantum Design). The Hall coefficient measurements were performed by the four-probe method with the PPMS under magnetic fields in the range from -7.0 to +7.0 T at intervals of 1.0 T.

## 3. Results and discussion
### 3.1 Characterization of samples
#### 3.1.1 As-synthesized sample

Figure 1(a) shows a powder XRD pattern for the as-synthesized sample together with a simulated XRD pattern for $K_8Al_8Si_{38}$. The parameters for $K_8Al_8Si_{38}$ reported in Ref. 23 were used in the simulation. All the observed reflections can be indexed as a cubic structure and are identical to the simulated pattern; therefore, it was concluded that the as-synthesized sample consists of a single phase Si clathrate. The chemical compositions of K, Al, and Si in the as-synthesized sample were determined using ICP-OES to be 19.4(4), 12.0(1), and 68.3(7) wt% (values in parentheses are the standard deviations), respectively. Thus, the chemical formula of the as-synthesized sample can be expressed by $K_{7.9(2)}Al_{7.1(1)}Si_{38.9(4)}$.

Figure 2 shows an NPD pattern of the as-synthesized sample together with the results of Rietveld analysis. The observed pattern can be reproduced well, assuming that the sample is the type-I Si clathrate $K_8Al_xSi_{46-x}$ with the parameters shown in Table I. No diffraction peak from an impurity phase was observed, which is consistent with the powder XRD pattern. The chemical formula derived using the Rietveld parameters is $K_8Al_{8.3(20)}Si_{37.6(20)}$, which is consistent with the ICP-OES results within the experimental error. The theoretical density was calculated to be 2.32(3) g/cm$^3$ using the lattice parameter and the chemical compositions determined from ICP-OES measurements.

The lattice parameter ($a$ = 10.442 Å) is consistent with that determined by powder XRD ($a$ = 10.434(1) Å) [25]; it is smaller than those of $K_8Al_8Si_{46}$ ($a$ = 10.4802(16) Å [23]) and $K_8Al_{23}Si_{23}$ ($a$ = 10.563 Å) [30] and much larger than that of $K_8Si_{46}$ ($a$ = 10.27518(5) Å) [31]. These observations are consistent with the increase in the lattice parameter in $BaAl_xSi_{46-x}$ with the Al content [32]. The values of x at the 16c site, y at the 24k site, and z at the 24k site are almost the same as those for $K_8Al_8Si_{38}$. The Al occupancy at the 6c site is much larger than that at the other two sites, 16i and 24k, and the Al occupancy at the 16i site is comparable to that at the 24k site within the experimental error. The high Al occupancy at the 6c site has also been observed in other ternary Si clathrates including $K_8Al_8Si_{38}$ [23]. The Al occupancies at the 16i and 24k sites of $K_8Al_7Si_{39}$ and $K_8Al_8Si_{38}$ are different. Therefore, further investigation is necessary to elucidate the difference in the Al occupancies of the 16c and 24k sites in these two clathrates.

### 3.1.2 SPS samples

Figure 1(b) shows an XRD pattern of the SPS sample. A small reflection from Si was observed (denoted as Si), which suggests that a small part of the sample was decomposed during SPS treatment.

The chemical composition of the SPS sample determined by EPMA was 18.8(2) wt. % K, 12.0(6) wt. % Al, and 69.3(4) wt. % Si; therefore, the chemical formula can be expressed as $K_{7.6(1)}Al_{7.0(4)}Si_{39.3(2)}$, which is almost the same as $K_8Al_7Si_{39}$. The chemical composition of the SPS sample is almost the same as that of the as-synthesized sample, despite the decomposition of a small amount of the sample during SPS treatment.

The density of the SPS sample was estimated to be 1.975 g/cm$^3$, which corresponds to 85.1% of the theoretical density obtained from the NPD and ICP-OES data (2.32(3) g/cm$^3$).

### 3.2 Transport properties

Figure 3 shows the electrical resistivity $\rho$ for K$_8$Al$_7$Si$_{39}$ as a function of temperature in the range from 10 to 320 K. K$_8$Al$_7$Si$_{39}$ exhibits metallic conduction: $\rho$ increases with the temperature.

Figure 4(a) shows the variation of the Hall coefficient $R_H$ with the temperature. The inset shows the transverse resistivity $\rho_{xy}$ measured at 260 K as a function of the magnetic field $\mu_0 H$. $R_H$ was estimated from the slope calculated by the least-squares fitting of $\rho_{xy}$ against $H$. $R_H$ has a negative value, which indicates that electrons are the dominant carriers. The magnitude of $R_H$ is almost independent of the temperature. Figure 4(b) shows the carrier density $n$ calculated using the equation $R_H = 1/ne$; $n$ is almost independent of the temperature. The temperature dependence of $\rho$ and $n$, and the magnitude of $n$ (ca. 1×10$^{27}$ m$^{-3}$) indicate that K$_8$Al$_7$Si$_{39}$ is a metal.

Figure 5 shows the temperature dependence of the Hall mobility $\mu$ calculated using the equation $\mu = R_H/\rho$. $\mu$ is independent of the temperature, which suggests that the carriers are scattered by neutral impurities in this temperature range [33].

Figure 6 shows the dependence of the Seebeck coefficient $S$ on the temperature in the range of 10-320 K. $S$ has negative values over the entire temperature range, which indicates that the electrons are dominant carriers. This is consistent with the results of Hall coefficient measurements. The absolute value of $S$ increases linearly with the temperature, which suggests that K$_8$Al$_7$Si$_{39}$ has no strong electron-phonon interaction.

The effective mass $m^*$ at 260 K was estimated using the values of $n$ and $S$ given by the following two equations:

$$S = \pm \left(\frac{k_B}{e}\right)\left\{\frac{(r+5/2)F_{r+3/2}(\xi)}{(r+3/2)F_r(\xi)} - \xi\right\}, \quad (1)$$

$$n = 4\pi \left(\frac{2m^* k_B T}{h^2}\right) F_{1/2}(\xi), \quad (2)$$

where $e$, $k_B$, $h$, $F_n(\xi)$, $r$, and $\xi$ respectively represent the charge of an electron, the Planck constant, the Boltzmann constant, the $n$th-order Fermi integral, a scattering parameter, and the reduced Fermi energy defined by $\xi = E_F/k_B T$, where $E_F$ is the Fermi energy. These equations were derived from the Boltzmann transport equation within the relaxation time approximation, assuming that electron conduction occurs within a single parabolic band and the energy dependence of the relaxation time is $E^{r+3/2}$ [34,35]. In these equations, the origin of $E_F$ is defined by the bottom of the conduction band. The value of $\xi$ was estimated to be 11.9 using the measured $S$ value and Eq. (1), assuming that the carriers are scattered by neutral impurities ($r=3/2$), as suggested by the temperature dependence of $\mu$. The $\xi$ value suggests that $K_8Al_7Si_{39}$ is a degenerated conductor, which is consistent with the fact that $K_8Al_7Si_{39}$ exhibited metallic conduction. The value of $m^*$ was calculated using the values of $\xi$ and $n$ with Eq. (2). The estimated $m^*$ value is tabulated in Table II together with the $m^*$ values of other ternary Si or Ge clathrates [23,36,37]. The $m^*$ value of $K_8Al_7Si_{39}$ is larger than that calculated for $K_8Al_8Si_{38}$ [23] and is comparable to those of $Ba_8Ga_xSi_{46-x}$ [36] and $Ba_8Ga_{16-x}Ge_{30+x}$ [37].

The total thermal conductivity $\kappa$, electronic thermal conductivity $\kappa_e$, and lattice thermal conductivity $\kappa_L$ are plotted as a function of temperature in Fig. 7. The inset exhibits the temperature dependence of $\kappa_L$ at temperatures ranging from 2 to 300 K on a log-log scale. $\kappa_e$ is estimated using the Wiedemann-Franz law, $\kappa_e = LT/\rho$, where $L$ is the Lorenz number. $\kappa_L$ was obtained by subtracting $\kappa_e$ from $\kappa$. The value of $L$ was estimated over the entire temperature range using the estimated $\xi$ value and

$$L = \left(\frac{k_B}{e}\right)^2 \left\{\frac{(r+7/2)F_{r+5/2}(\xi)}{(r+3/2)F_r(\xi)} - \left(\frac{(r+5/2)F_{r+3/2}(\xi)}{(r+3/2)F_r(\xi)}\right)^2\right\}. \quad (3)$$

The value of $L$ is almost independent of the temperature ($2.44 \times 10^{-8}$ V$^2$K$^{-2}$ at 11 K and $2.43 \times 10^{-8}$ V$^2$/K$^{-2}$ at 300 K), which is almost the same as the $L$ value evaluated for a degenerated metal, $L_0 = (\pi^2/3)(k_B/e)^2$ ($2.44 \times 10^{-8}$ V$^2$/K$^{-2}$). The electronic contribution of the thermal conductivity $\kappa_e$ of K$_8$Al$_7$Si$_{39}$ is lower than that of the other clathrates owing to its higher $\rho$ value. The temperature dependence of $\kappa_L$ is similar to that of a crystalline solid: $\kappa_L$ increases steeply with the temperature up to a maximum at approximately 50 K without showing a plateau, then decreases gradually. $\kappa_L$ at 300 K is 1.7(2) WK$^{-1}$m$^{-1}$, which is comparable to the calculated value reported by He and Galli [24].

The dimensionless figure of merit $ZT$ is plotted as a function of temperature in Fig. 8. The $ZT$ value is 0.0096 at 300 K. Table III summarizes $\rho$, $S$, $\kappa$, and $ZT$ for K$_8$Al$_7$Si$_{39}$ at 300 K, together with those for other Si or Ge clathrates [14, 17, 38-41]. The values of $\rho$, S, $\kappa$, and $ZT$ for semiconducting K$_8$Al$_8$Si$_{38}$, reported very recently [38], are also listed in Table III. The values for K$_8$Al$_7$Si$_{39}$ and K$_8$Al$_8$Si$_{38}$ were compared initially. The $\rho$ and $S$ values of K$_8$Al$_7$Si$_{39}$ are much smaller than those of K$_8$Al$_8$Si$_{38}$, which is consistent with the metallic nature of K$_8$Al$_7$Si$_{39}$, while K$_8$Al$_8$Si$_{38}$ is semiconducting. However, the $\kappa$ values are comparable. The $ZT$ value for K$_8$Al$_7$Si$_{39}$ is approximately five times larger than that for K$_8$Al$_8$Si$_{38}$. Subsequently, we compare the values of K$_8$Al$_7$Si$_{39}$ and those of clathrates other than K$_8$Al$_8$Si$_{38}$. The $\rho$ value for K$_8$Al$_7$Si$_{39}$ is several times higher than that for the other clathrates. $S$ and $\kappa$ for K$_8$Al$_7$Si$_{39}$ are comparable with those for the other clathrates. The *ZT* value for $K_8Al_7Si_{39}$ is comparable with that for the Si clathrates, except for $Ba_8Ga_{16}Si_{30}$, and is approximately one-fifth of those for $Ba_8Ga_{16}Si_{30}$ and $Ba_8Ga_{16}Ge_{30}$ at 300 K. It should be noted that the *ZT* value for $Ba_8Ga_{16}Ge_{30}$ increases with the temperature and reaches 0.80 at 1000 K [14]. The high $\rho$ value leads to the lower *ZT* value in $K_8Al_7Si_{39}$; therefore, a higher *ZT* value could be achieved if $\rho$ could be decreased without significant influence on the other parameters. Furthermore, the calculated *ZT* for $K_8Al_8Si_{38}$ is approximately 0.3 at 300 K and 0.6 at 900 K, assuming $\kappa_L$ of 1 $WK^{-1}m^{-1}$ [42]. Therefore, the optimization of the SPS conditions is necessary to decrease $\rho$ and lead to an improvement of *ZT*. Efforts toward the optimization of these conditions are under way.

The present results indicate that $K_8Al_7Si_{39}$ is a metal with electrons as dominant carriers, while $K_8Al_8Si_{38}$ has been reported to be a semiconductor with an energy gap of approximately 1.0 eV [23]. We consider that the metallic nature of $K_8Al_7Si_{39}$ can be attributed to the deviation of the chemical composition from its stoichiometry. $K_8Al_xSi_{46-x}$ is a Zintl phase [4], in which K atoms donate their valence electrons to Al atoms, and Al atoms with an excess valence electron form a polyhedral network with Si atoms. When the Al content $x$, is eight, the valence electrons from eight K atoms are completely compensated by eight Al atoms. As a result, the Fermi energy $E_F$ is located within $E_g$ and $K_8Al_8Si_{38}$ becomes a semiconductor. This was clearly explained in the case of $K_8Ga_8Si_{38}$ using the density of states of $Si_{46}$, $K_8Si_{46}$, and $K_8Ga_8Si_{38}$ [19]. $K_8Al_8Si_{38}$ has been demonstrated to be a semiconductor, both computationally and experimentally [20, 23, 38, 42, 43]. On the other hand, the chemical formula of the clathrate synthesized in the present study is $K_8Al_7Si_{39}$, which has two more valence electrons than $K_8Al_8Si_{38}$; therefore, $E_F$ for $K_8Al_7Si_{39}$ shifts to a higher energy than $E_F$ for $K_8Al_8Si_{38}$, and is located in a state that corresponds to the conduction band of $K_8Al_8Si_{38}$. As a result, $K_8Al_7Si_{39}$ becomes a metal.

## 4. Conclusions

A ternary Si clathrate, $K_8Al_7Si_{39}$, was synthesized and its transport properties were examined at temperatures in the range from 10 to 320 K. Measurements of $\rho$ and $R_H$ revealed that $K_8Al_7Si_{39}$ is a metal with electrons as the dominant carriers at a density of approximately $1\times10^{27}$ m$^{-3}$ and with $\mu$ of $1\times10^{-4}$ m$^2$V$^{-1}$s$^{-1}$. $S$ is negative and its absolute value increases with the temperature. The temperature dependence of $\kappa$ is similar to that of a crystalline solid. The $ZT$ value is approximately 0.01 at 300 K, which is better than that for $K_8Al_8Si_{38}$ and is comparable to that for other ternary Si clathrates. Further investigation of the transport properties at higher temperatures using samples with various Al concentrations will be important for clarifying the potential of $K_8Al_xSi_{46-x}$ as a functional material.

**Acknowledgements**

The authors are thankful to Dr. T. Suzuki and Dr. Y. Sakka of National Institute for Materials Science (NIMS) for allowing the use of SPS apparatus, Dr. M. Nishio of NIMS for EPMA and A. Ishitoya of NIMS for ICP-OES. We also thank Dr. Y. Matsushita of NIMS for advice on the handling of potassium. Dr. Shinohara and Dr. Tsujii of NIMS are acknowledged for valuable discussion on thermoelectric properties. This research was partially supported by the ALCA program of the Japan Science and Technology Agency (JST).

Table I. Crystal structure of $K_8Al_xSi_{46-x}$. Cubic. Space group: $P\bar{m}3n$ (No. 223). Lattice parameters: $a = 10.442$ Å. $Z = 1$. $R$-factors (%): $R_p = 11.9$, $R_{wp} = 10.4$, $R_e = 8.48$, and $\chi^2 = 1.520$. The definitions of $R_p$, $R_{wp}$, $R_e$, and $\chi^2$ are given in Ref. 27. The values in parentheses are the standard deviations. $B_{iso}$ represents the equivalent isotropic displacement parameter.

| Label | Species | Multiplicity Wyckoff Lett. | $x$ | $y$ | $z$ | $B_{iso}$ (Å$^2$) | Occupancy |
|---|---|---|---|---|---|---|---|
| K1 | K | 2a | 0.00000 | 0.00000 | 0.00000 | 1.16(7) | 1.000 |
| K2 | K | 6d | 0.25000 | 0.50000 | 0.00000 | 2.08(7) | 1.000 |
| Si1/Al1 | Si/Al | 6c | 0.25000 | 0.00000 | 0.50000 | 0.84(5) | 0.15(6) /0.85(6) |
| Si2/Al2 | Si/Al | 16i | 0.18440(5) | x | x | 0.78(2) | 0.89(4) /0.11(4) |
| Si3/Al3 | Si/Al | 24k | 0.00000 | 0.30263(7) | 0.11664(8) | 0.84(2) | 0.94(4) /0.06(4) |

**Table II.** Effective mass $m^*$ for $K_8Al_7Si_{39}$ together with those for other ternary Si or Ge clathrates [23, 36, 37], where $m_e$ represents the electron rest mass.

| Material | $m^*$ ($m_e$) | Reference |
|---|:---:|:---:|
| $K_8Al_7Si_{39}$ | 1.46 | Present |
| $K_8Al_8Si_{38}$ (calc) | 0.68 for electrons | 23 |
|  | 1.03 for holes |  |
| $Ba_8Ga_xSi_{46-x}$ | 1.5 | 36 |
| $Ba_8Au_xGa_ySi_{46-x-y}$ | 2.0 | 36 |
| $Ba_8Sr_2AuGa_{13}Si_{32}$ | 3.0 | 36 |
| $Ba_8Ga_{16-x}Ge_{30+x}$ | 1.82 | 37 |

**Table III.** Electrical resistivity $\rho$, Seebeck coefficient $S$, thermal conductivity $\kappa$, and dimensionless figure of merit $ZT$ for $K_8Al_7Si_{39}$ at 300 K together with those for other Si or Ge clathrates [14, 17, 38-41].

| Material | $\rho$ (μΩ m) | $S$ (μVK$^{-1}$) | $\kappa$ (WK$^{-1}$m$^{-1}$) | $ZT$ | Reference |
|---|---|---|---|---|---|
| $K_8Al_7Si_{39}$ | 78.5(6) | -65.5(5) | 1.7(2) | 0.0096(15) | Present |
| $K_8Al_8Si_{38}$ | 6.4×10$^3$ | -222 | 1.65 | 0.0014 | 38 |
| $Ba_8Al_{14}Si_{31}$ | 5.8 | -27 | 2.8 | 0.014 | 39 |
| $Sr_{0.7}Ba_{7.3}Al_{14}Si_{31}$ | 14 | -32 | 1.5 | 0.015 | 40 |
| $Ba_8Ga_{16}Si_{30}$ | 20 | -66 | 1.2 | 0.055 | 17 |
| $Ba_8Ni_{3.8}Si_{42.2}$ | 4.0 | -26.1 | 4.2 | 0.012 | 41 |
| $Ba_8Ga_{16}Ge_{30}$ | 6.4 | -45 | 1.75 | 0.054 | 14 |

Figure Captions

Fig. 1. Powder XRD patterns for (a) as-synthesized and (b) SPS samples together with a simulated pattern for $K_8Al_8Si_{38}$.

Fig. 2. NPD pattern for as-synthesized $K_8Al_7Si_{39}$. The red bars and black line represent the observed and calculated intensities, respectively. The difference between the two intensities is indicated by the blue line. Peak positions for $K_8Al_7Si_{39}$ are labeled with green vertical bars.

Fig. 3. Electrical resistivity $\rho$ for $K_8Al_7Si_{39}$ as a function of temperature $T$.

Fig. 4. (a) Hall coefficient $R_H$ and (b) carrier density $n$ for $K_8Al_7Si_{39}$ as a function of temperature $T$. The inset shows the transverse resistivity $\rho_{xy}$ at 260 K as a function of the applied magnetic field $\mu_0 H$.

Fig. 5. Hall mobility $\mu$ for $K_8Al_7Si_{39}$ as a function of temperature $T$.

Fig. 6. Seebeck coefficient $S$ for $K_8Al_7Si_{39}$ as a function of temperature $T$.

Fig.7. Total thermal conductivity $\kappa$, electronic thermal conductivity $\kappa_e$, and lattice thermal conductivity $\kappa_L$ for $K_8Al_7Si_{39}$ as a function of temperature $T$. The inset shows the temperature dependence of $\kappa_L$ at temperatures from 2 to 300 K on a log-log scale.

Fig. 8. Dimensionless figure of merit $ZT$ for $K_8Al_7Si_{39}$ as a function of temperature $T$.

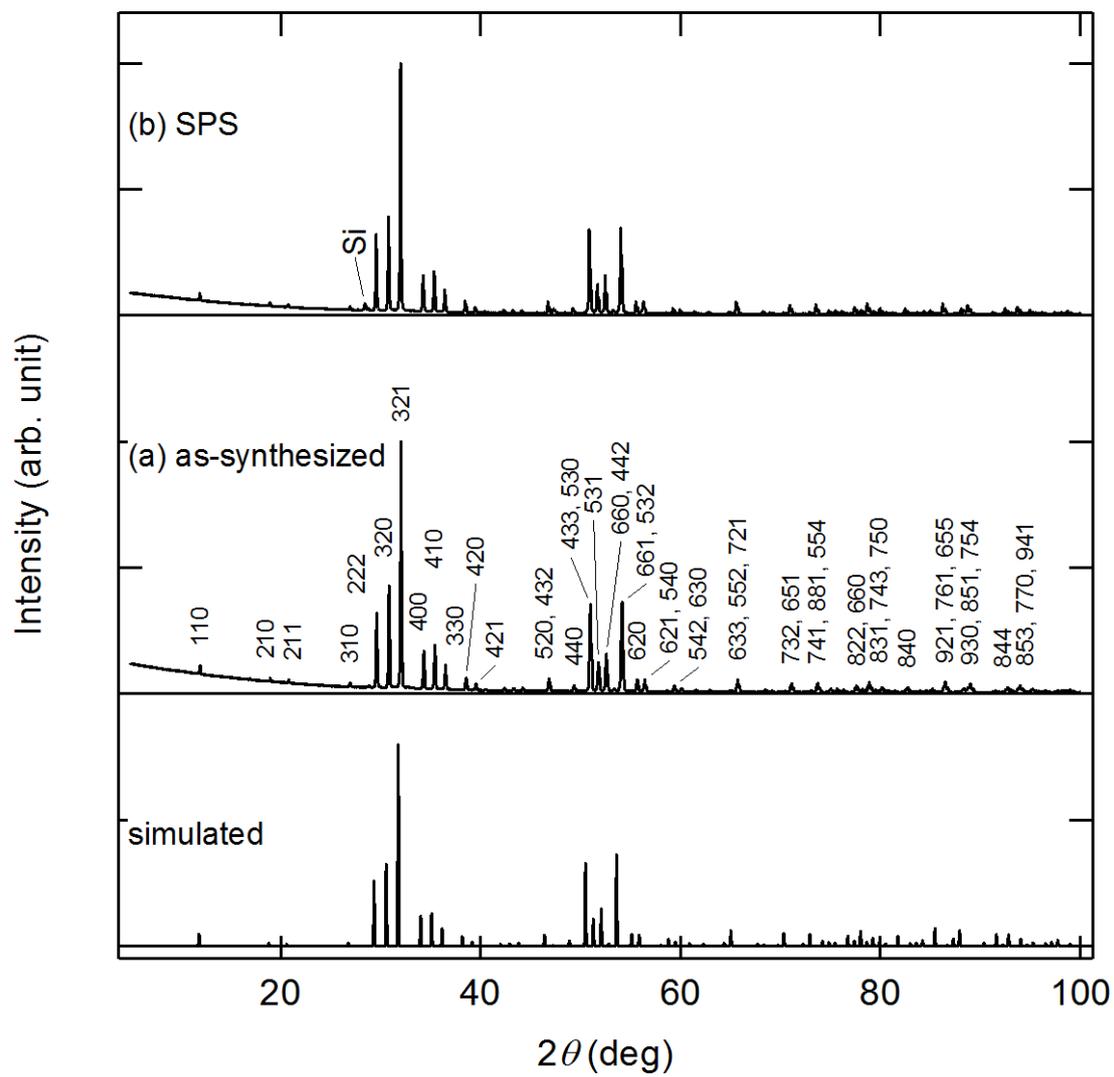

Fig. 1

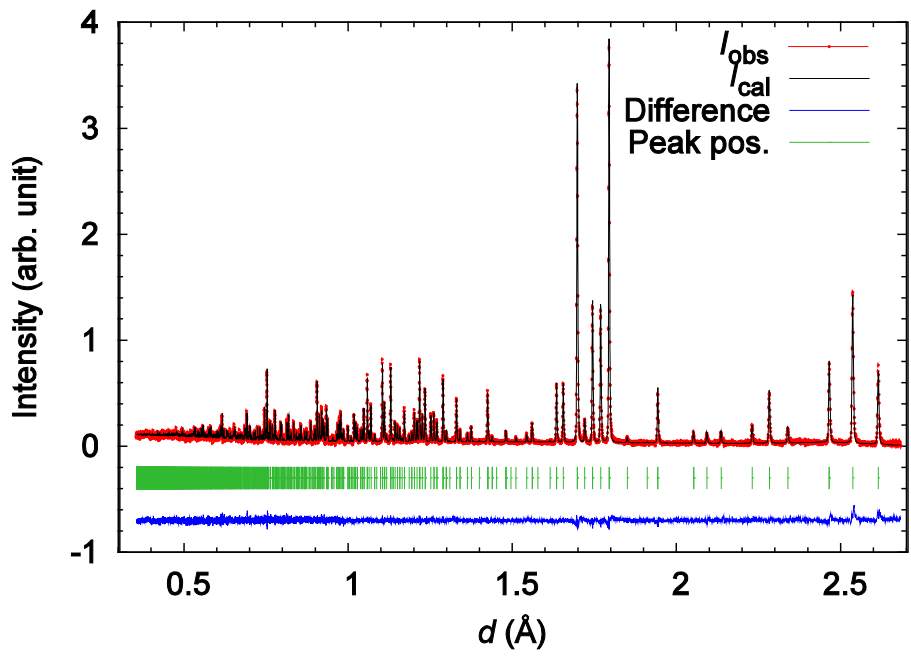

Fig. 2

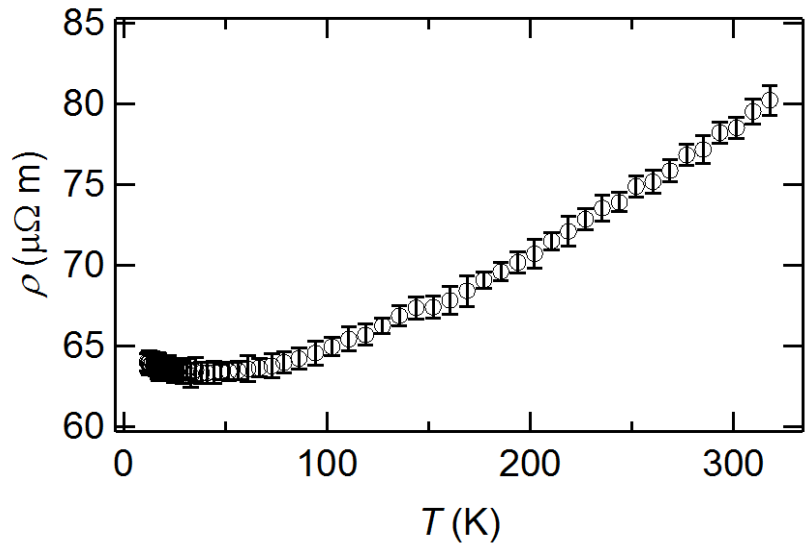

Fig. 3

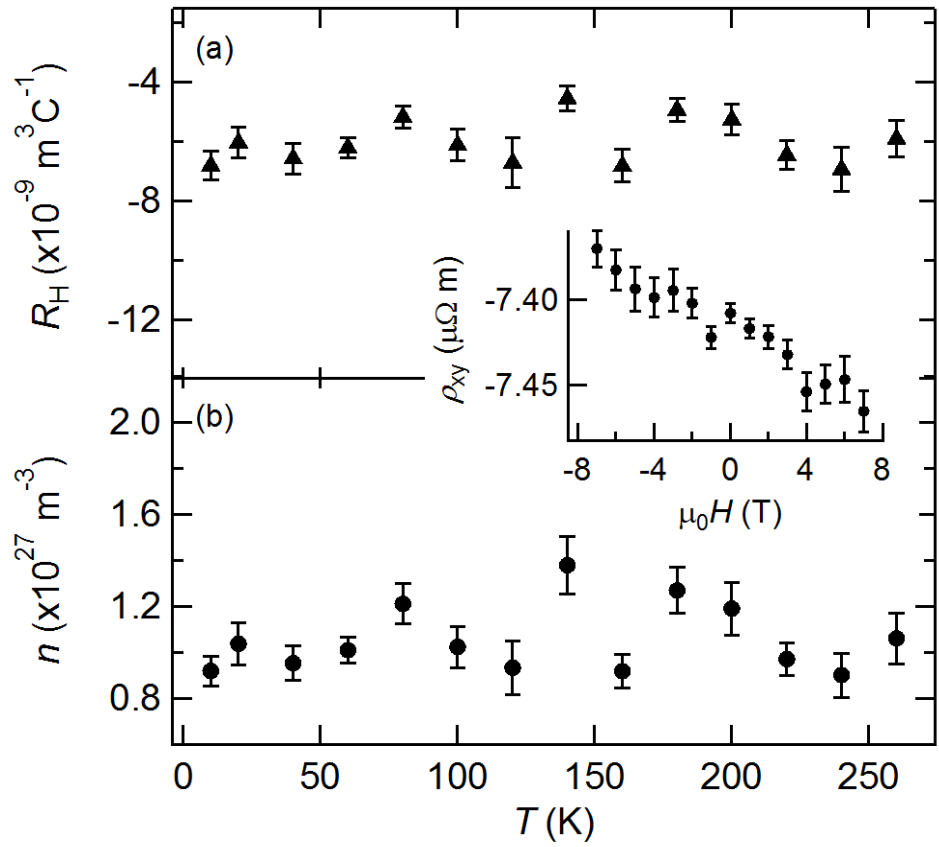

Fig. 4

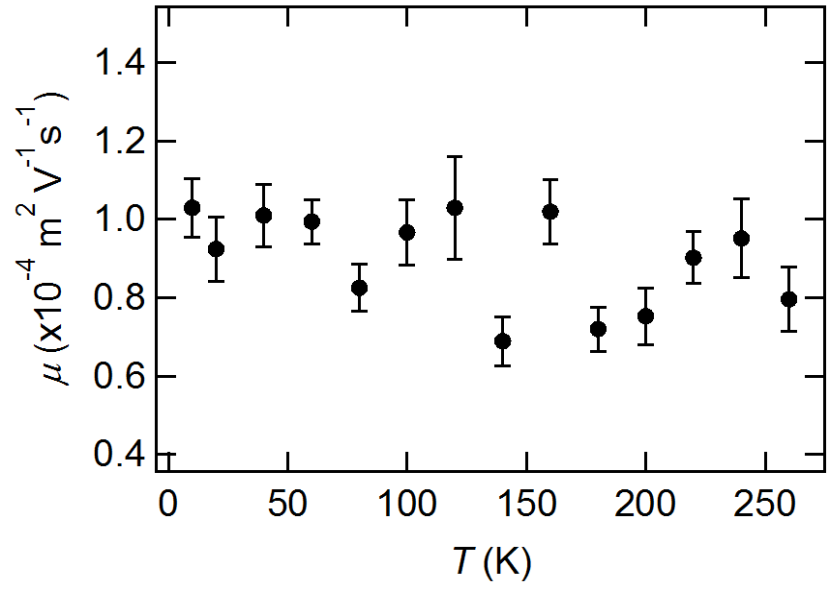

Fig. 5

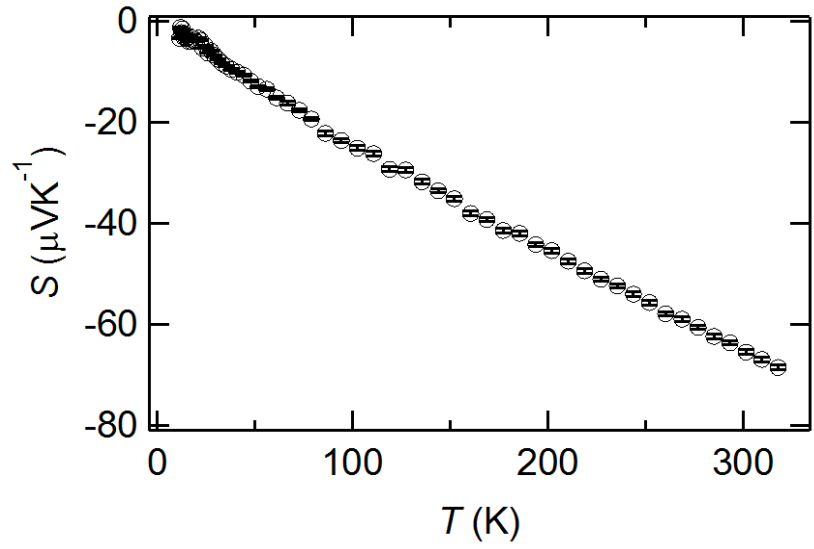

Fig. 6

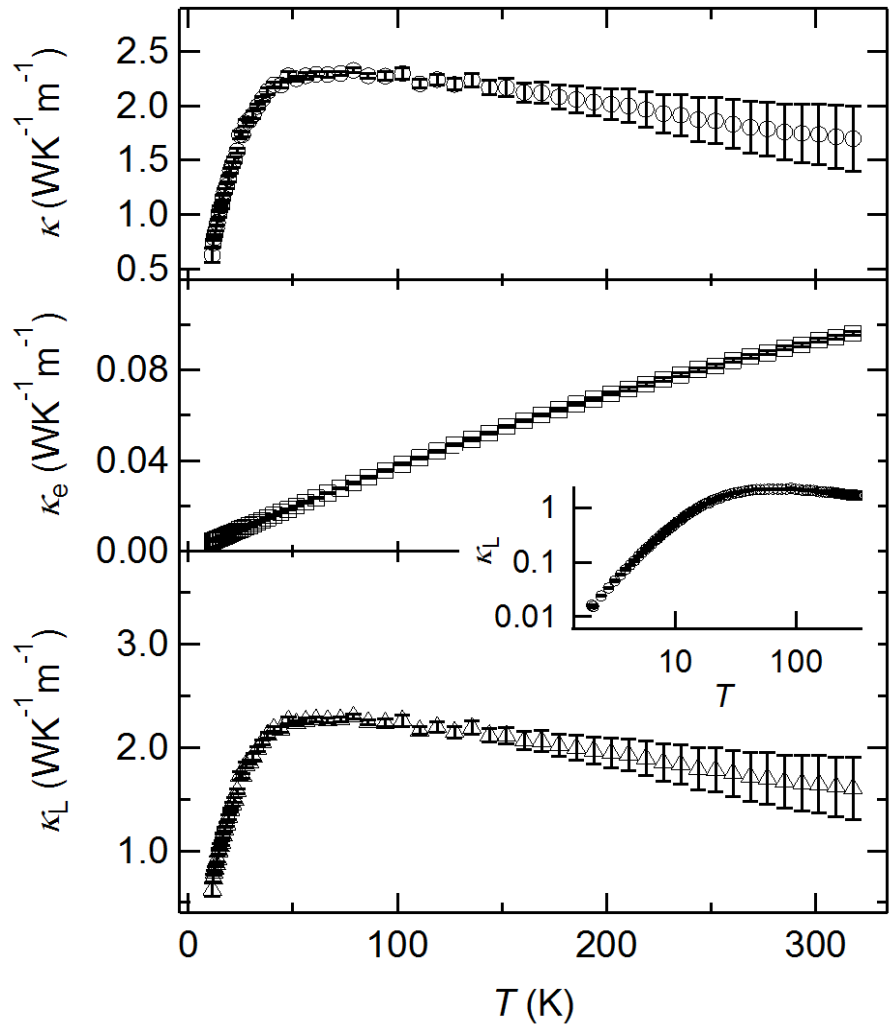

Fig. 7

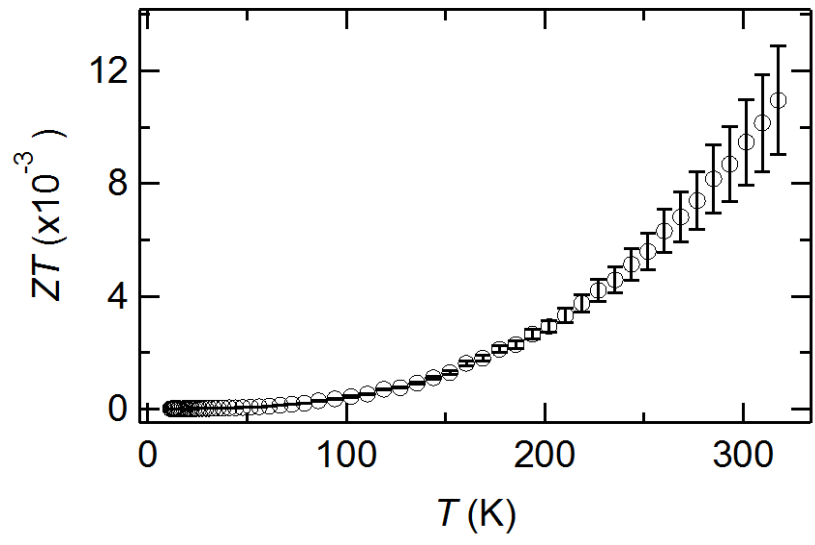

Fig. 8